\documentstyle[preprint,aps]{revtex}

\newcommand{\nl}{\nonumber \\}

\newcommand{\be}{\begin{equation}}
\newcommand{\ee}{\end{equation}}
\newcommand{\bea}{\begin{eqnarray}}
\newcommand{\eea}{\end{eqnarray}}

\newcommand{\Eq}[1]{Eq.\,(\ref{#1})}

\newcommand{\la}{\langle}
\newcommand{\ra}{\rangle}

\begin{document}
\draft
%\tighten

\title{ Electrical transport through individual DNA molecules }

\author{\bf Xin-Qi Li and YiJing Yan}

\address{Department of Chemistry, Hong Kong University of Science and 
         Technology, Kowloon,\\ Hong Kong }

%\date{\today}
\date{March 30, 2001}

\maketitle
\begin{abstract}
A theoretical study is presented to quantitatively analyze 
the transport experiment through individual DNA molecules
reported recently by Porath {\it et al.} [Nature {\bf 403}, 635 (2000)].
%%%%
A variety of valuable quantities are identified by contacting 
the theoretical model with the measured data. 
The partially decoherent nature on the GC pairs of DNA is elaborated 
in contrast to the completely incoherent hopping mechanism 
discussed in the context of charge transfer experiments.
\end{abstract}

\vspace{6ex}

%%%%%%%%%%%%%%%%%%%%%%%%%%%
%% \section{Introduction}
%%%%%%%%%%%%%%%%%%%%%%%%%%%

DNA has the special double-helix structure with 
$\pi$-electron cores of well stacking bases, which may provide 
an one-dimensional pathway for charge transport.
Since first posed in 1960s, the question whether or not DNA is able 
to conduct electrical charges is still debated.
Due to the advent of molecular electronics as well as other 
biological considerations,
this issue has stimulated intense research interest.
Recent studies proposed that DNA seems to be one of the most promising 
candidates in the application of functional nano-electronic devices
\cite{Mir96607}.
%%%%
%%%%
Very recently, direct measurements for the electrical transport 
through DNA molecules were performed \cite{Fin99407,Por00635}. 
In particular, Porath {\it et al.} \cite{Por00635} measured the electrical
transport through the individual DNA molecules.
The observed nonlinear current-voltage (I-V)
characteristics clearly suggest
that DNA molecules be good molecular semiconductors.
%%%%%%%%%
In this letter, we carry out a quantitative analysis for the experimental
I-V characteristics \cite{Por00635}, and discuss the relevant physical
implications.

%%%%%%%%%%%%%%%%%%
%% \section{Theoretical model}
%%%%%%%%%%%%%%%%%%

The theoretical model for the experimental setup is schematically shown
in Fig.\ 1. 
In the experiment \cite{Por00635}, the DNA sample of 10.4 nm-long chain
containing 30 GC pairs was studied.
%%%%
The {\it ab initio} calculation \cite{Zha01}
showed a large HOMO-LUMO gap (10.07 eV) of the single GC pair.
We thus adopt a homogeneous one-band tight-binding model
for the HOMO-mediated charge transport 
through the DNA molecules under study.
%%%%%
To proceed, 
consider the following Hamiltonian to describe the system in Fig.\ 1: 
\bea
H   &=& H_0 + H_I ,    \nl
H_0 &=& H_M + H_L + H_R + H_{\rm res} .
\eea
Here $H_0$ describes the separated subsystems of the DNA molecule ($H_M$),
the left and right metal electrodes ($H_L$ and $H_R$), and the 
dephasing reservoirs ($H_{\rm res}$). 
$H_I$ couples the DNA molecule to the electrodes and the dephasing
reservoirs.
%%%%%%
Following B\"uttiker's idea for the inelastic scattering effect
\cite{But863020,Dam907411}, 
the phase-breaking processes on the GC-pairs are modeled 
by coupling each GC-pair
to a {\it fictitious electronic reservoir}, see Fig.\ 1.
The dephasing strength is characterized by the coupling parameter $\eta$.

To carry out the nonlinear transport characteristics, the knowledge of 
the energy-dependent transmission coefficient through the molecule 
is required.
Owing to the open boundary conditions associated with the voltage electrodes 
and dephasing reservoirs, the Green's function method would be the 
proper and versatile theoretical tool.
We thus introduce the following Green's function
\bea
{\cal G}(E)=(E-{\cal H}_{\rm eff})^{-1} ,
\eea
where $\cal{H}_{\rm eff}$ is the effective Hamiltonian of the reduced 
molecular system, obtained by eliminating the degrees of freedom of 
the two electrodes and the dephasing reservoirs.
%%% {\bf (2)}
As a simple model, 
the electrodes and dephasing reservoirs can be described by semi-infinite 
one-dimensional tight-binding chains \cite{Dam907411}, 
which result in the self-energy
corrections to the Hamiltonian of the molecule as 
\bea
{\cal H}_{\rm eff}
    =  H_M + \Sigma_L |1\ra\la 1| +\Sigma_R |N\ra\la N|
       + \sum_{j=1}^{N} \Sigma_j |j\ra\la j|  .
\eea
$\Sigma_{L(R)}$ and $\Sigma_j$ are respectively the self-energies
resulting from the coupling of the molecule to
the left (right) electrodes, and the $j$th dephasing reservoir.
Using Dyson equation, 
the self-energy $\Sigma_{\mu}$ ($\mu=L, R, {\rm and} \{j\}$)
can be carried out as
\bea
\Sigma_{\mu}&=& \frac{V_{\mu}^2 }{E-E_{\mu}-\sigma_{\mu}} ,
\eea
where $\sigma_{\mu}$ is the self-energy correction of the semi-infinite chain
to the ending-site of the chain
attached to the DNA molecule, and has the following
expression \cite{Dam907411} 
\bea
\sigma_{\mu} = \frac{E-E_{\mu}}{2}
-i\left[\gamma_{\mu}^2-\left(\frac{E-E_{\mu}}{2}\right)^2 \right]^{1/2} .
\eea
Here $V_{\mu}$ are the coupling strengths
between the molecule and the electrodes/reservoirs: 
$V_{\mu}=V_{L(R)}$ if $\mu=L(R)$; 
and $V_{\mu}=\eta$ if $\mu=\{j\}$.
$E_{\mu}$ and $\gamma_{\mu}$ denote the site energies and the nearest-neighbor
hopping strengths of the homogeneous tight-binding electrode and reservoir 
chains. 
We let $\gamma_{L(R)}=\gamma$ for the left (right) electrode and 
$\gamma_{j}=\gamma_0$ for the dephasing reservoirs.
Thus, $4\gamma$ and $4\gamma_0$ characterize the band widths of the
electrodes and the dephasing reservoirs,
which are assumed in this work to be commonly 5 eV.

To connect the Green's function with the transmission coefficients between 
any pair of reservoirs, say, from $\mu$ to $\nu$,
consider a linear transport between them. 
%%%%%
It can be shown \cite{Muj946849}
that the linear conductance at low temperatures is 
$g_{\mu\nu}=(2e^2/\pi\hbar)\Delta_{\mu}\Delta_{\nu} |{\cal G}_{\mu\nu}(E)|^2$.
Here, $\Delta_{\mu}$ is associated with the self-energy
by $\Delta_{\mu}=-{\rm Im}\Sigma_{\mu}$.
${\cal G}_{\mu\nu}(E)$ is the matrix element of ${\cal G}(E)$ between the
molecular sites connecting with the $\mu$th and $\nu$th reservoirs 
($\mu,\nu=L, R, {\rm and} \{j\}$).
%%%%%%
On the other hand, the linear conductance is given by the Landauer-B\"uttiker
formula \cite{Dat95}
as $g_{\mu\nu}=(e^2/h) T_{\mu\nu}$. 
We therefore establish the following desired relationship 
between the transmission coefficient $T_{\mu\nu}$ and the Green's function:
\bea \label{Tmn}
T_{\mu\nu} = 4 \Delta_{\mu}\Delta_{\nu} |{\cal G}_{\mu\nu}(E)|^2  .
\eea
%%%%%%%%%%%
%% Here we have adopted the well-known wide-band limit approximation for 
%% the electrodes and reservoirs, which leads to
%% $\Sigma_{\mu}\simeq -i\Delta_{\mu}$.
%% For the left (right) electrode, $\Delta_{\mu}=V^2_{L(R)}/\gamma$; 
%% and for the dephasing reservoirs, $\Delta_{\mu}=\eta/\gamma_0$.
%%%%%%%%%%%
Note that in the wide band limit, $\Delta_{\mu}\simeq V^2_{\mu}/\gamma_{\mu}$.
We thus conclude that in the special case $V_{\mu}=\gamma_{\mu}$,
\Eq{Tmn} reduces to the result of Refs.\ \cite{Dam907411,Fis816851}.
%%%%%
%%%%%
After knowing $T_{\mu\nu}$, we are ready to carry out the total 
{\it effective transmission coefficient} from the left electrode 
to the right one.
We have recently shown 
that for each individual electron 
being emitted from the electrode, the effective transmission coefficient
in the presence of phase-breaking is of the same form as that first derived
by D'Amato and Pastawski \cite{Li01,Dam907411}: 
\bea \label{Teff}
{\cal T}_{\rm eff}(E)
    = T_{LR}+\sum^{N}_{\mu,\nu=1} K^{(L)}_{\mu}
      W^{-1}_{\mu\nu} K^{(R)}_{\nu} .
\eea
Here $K^{(L)}_{\mu}=T_{L\mu}$, $K^{(R)}_{\nu}=T_{\nu R}$.
$W^{-1}$ is the inverse of the matrix $W$ defined in terms of 
the matrix elements
$W_{\mu\nu}=(1-R_{\nu\nu})\delta_{\mu\nu}-T_{\mu\nu}(1-\delta_{\mu\nu})$, 
with $R_{\nu\nu}=1-\sum_{\mu(\neq \nu)}T_{\nu\mu}$.
The physical meaning of \Eq{Teff} is clear:
the first term is the coherent contribution of tunneling, whereas 
the second term denotes the incoherent component
owing to electron suffering the dephasing processes.

With the knowledge of the transmission function ${\cal T}_{\rm eff}(E)$,
it is straightforward 
to evaluate the I-V characteristics by applying the standard
formalism based on the scattering theory of transport \cite{Dat95}
\bea \label{IV}
I = \frac{2e}{h}\int_{-\infty}^{\infty} dE 
     {\cal T}_{\rm eff}(E) [f_L(E)-f_R(E)] .
\eea
Here $f_{L/R}(E)=\{{\rm exp}[(E-\mu_{L/R})/k_{\rm B}T] +1\}^{-1}$
is the Fermi function. The room temperature ($k_{\rm B}T \simeq 26$ meV)
will be considered in the numerical results, in order to keep 
consistency with the experiment \cite{Por00635}.
$\mu_{L}$ and $\mu_{R}$ are the electrochemical potentials of the two 
metal electrodes, whose values depend on the applied bias voltage.
%%%%%
Following Datta {\it et al.} \cite{Dat972530}, we set
$\mu_L=E_f+(1-\kappa) e{\cal V}$, and 
$\mu_R=E_f-\kappa e{\cal V}$, 
where $E_f$ and ${\cal V}$ are respectively the equilibrium Fermi energy
and the applied voltage.
%%%%%
$\kappa$ is a useful parameter in characterizing 
how the applied voltage ${\cal V}$
is divided across the two junctions between the molecule and the electrodes.

%%%%%%%%%%%%%%%%%%%%%%%%%%%%%%%%%
%%%% \section{Numerical results}
%%%%%%%%%%%%%%%%%%%%%%%%%%%%%%%%%

Figure 2 shows the calculated I-V characteristics, 
together with the experimental data taken from Ref.\ \cite{Por00635}.
Below we discuss how to make contact of the theoretical model with 
the measured data in order to get useful informations.
%% {\it (i)} 
Firstly, we notice the asymmetric gaps $V_c^{(\pm)}$
under positive and negative bias voltages:
$V_c^{(+)}=1.6$ eV and $V_c^{(-)}=0.8$ eV.
This asymmetry suggests $\kappa=1/3$, 
instead of $1/2$ as usual \cite{Dat972530}.
%% {\it (ii)}
Secondly,
the experimental differential conductance \cite{Por00635} showed a
peak structure with the voltage spacings $\Delta {\cal V}$ of
$0.1\sim  0.5$ eV.
By taking the average value $\Delta {\cal V}\simeq 0.25$ eV
and $\kappa=1/3$,  we estimate the DNA molecular energy
level spacing as $\Delta E=\kappa\Delta {\cal V}\simeq 0.08$ eV.
Since the DNA under study consists of 30 GC-pairs, 
from the simple consideration
that the energy bandwidth $4t=30\times 0.08=2.4$ eV, 
we obtain the important information for  
the hoping strength between the nearest-neighbor GC pairs, 
$t\simeq 0.6$ eV, 
which is in good agreement with the {\it ab initio} calculation \cite{Zha01}.
%% {\it (iii)}
Thirdly, the equilibrium Fermi level $E_F$ of the electrodes
can be estimated via $E_F-E_G=2t+\kappa V_c^{(+)}\simeq 1.73$ eV,
where $E_G$ is the HOMO level of the single GC pair.
This information would be helpful in understanding and predicting
the voltage gaps in transport experiments.
%% {\it (iv)}
Finally, we notice the asymmetry of the entire profile of the experimental 
I-V curve under positive and negative bias voltages.
We temporarily understand this feature by the distinct coupling strengths 
of the DNA molecule to the metal electrodes:
$V_{L/R}=3$ meV for positive voltage, and
$V_{L/R}=2$ meV for negative voltage.
With these parameters, the theoretical results are in quantitative
agreement with the measured data as shown in Figure 2.

Since the I-V characteristic is an integrated result of the transmission
coefficient, the dephasing effect is not quite prominent on the I-V curve,
except that the voltage gaps and the current steps 
due to the discrete molecular energy levels 
are smoothed by the phase-breaking scattering as shown in Fig.\ 2. 
%%%%%%
However, it has drastic effect on the differential conductance 
as shown in Fig.\ 3.
Roughly speaking, the dephasing would broaden the conductance peaks.
More precisely, the differential conductance $dI/dV$ which corresponds
to the transmission coefficient ${\cal T}_{\rm eff}(E)$ at the specific voltage 
would change with the 
dephasing strength as representatively shown in the inset of Fig.\ 3.
The turnover behavior has also been discovered in other context of
tunneling with dissipation \cite{Li01}.
%%%%%%%
Note that the experimental peak width is of $\sim 0.2$ eV,
while its spacing $\sim 0.25$ eV.
It may thus suggest that the charge motion along the 
stacked GC-pairs is only partially dephased in the referred 
experiment \cite{Por00635}.
It is worth emphasizing
that this motional nature is essentially different from  
the hopping mechanism discussed in the context of long-range 
charge transfer phenomena in DNA \cite{Meg9812950}, where the 
quantum coherence is regarded as being completely destroyed on the GC pairs.
We may attribute this distinct nature to the different environments
in which the DNA molecules located.
%%%%%%%%%
Since the electronic dephasing would significantly influence
the long-range charge transfer efficiency \cite{Li01}, 
we here suggest to further identify the 
dephasing strength on the GC pairs, by investigating 
the electrical transport through various sequence of DNA molecules
and under different environments.

In summary, we have presented a theoretical formalism for the transport 
through DNA molecules. By making contact of the theoretical model with 
the experimental data, a variety of valuable quantities were identified,
and the theoretical result was in quantitative agreement with experiment.
The modest dephasing nature on the GC pairs was emphasized in particular,
viewing its close relevance to the long-range charge transfer in DNA.
Finally, the present theoretical formalism is quite general, 
which can also be employed to describe the transport through 
other molecular wires.

\section*{Acknowledgments}
    Support from the Research Grants Council of the Hong Kong Government
and the National Natural Science Foundation of China is gratefully
acknowledged.

% ======================================================================

% ==============================================================
\begin{figure}\label{Fig1}
\caption{ 
Schematic diagram for the electrical transport
through DNA molecules. The DNA molecular wire consists of a stack of GC
pairs, and each GC pair is connected with a fictitious electronic 
dephasing reservoir by a coupling strength $\eta$.  }
\end{figure}

\begin{figure}\label{Fig2}
\caption{I-V characteristics: theoretical results versus 
experimental data taken from Ref.\ [3].
The transport currents in the presence of
weak dephasing ($\eta=0.05$ eV) and stronger one ($\eta=0.3$ eV)
are theoretically shown by the solid and dashesd curves.  }
\end{figure}

\begin{figure}\label{Fig3}
\caption{ Dephasing effect on the differential conductance. 
The solid and dashed curves correspond to $\eta=0.05$ eV and 
$\eta=0.3$ eV, respectively.
The detailed change of the conductance profile is representatively 
shown in the inset. }
\end{figure}

% ==============================================================

\end{document}